\documentclass[english,aps,prd,twocolumn,superscriptaddress,preprintnumbers,nofootinbib]{revtex4}
\usepackage[T1]{fontenc}
\usepackage[latin9]{inputenc}
\usepackage{lmodern}
\usepackage{verbatim}
\usepackage{amsmath}
\usepackage{dcolumn}
\usepackage{graphicx}
\newcommand{\be}{\begin{equation}}
\newcommand{\ee}{\end{equation}}
\newcommand{\bea}{\begin{eqnarray}}
\newcommand{\eea}{\end{eqnarray}}

\usepackage{babel}
\begin{document}
\global\long\def\order#1{\mathcal{O}\left(#1\right)}
\global\long\def\d{\mathrm{d}}
\global\long\def\P{P}
\global\long\def\amp{{\mathcal M}}
\preprint{Alberta Thy 4-14}

\title{Muon decay spin asymmetry}

\author{Fabrizio Caola} 
\email{caola@pha.jhu.edu}
\affiliation{Department of Physics and Astronomy, Johns Hopkins 
University, Baltimore, USA}
\author{Andrzej Czarnecki} 
\email{andrzejc@ualberta.ca}
\affiliation{Department of Physics, University of Alberta, Edmonton, Canada}
\author{Yi Liang}
\email{alavan.yi@gmail.com}
\affiliation{Department of Physics, University of Alberta, 
Edmonton, Canada}
\author{Kirill Melnikov}
\email{melnikov@pha.jhu.edu}
\affiliation{Department of Physics and Astronomy, Johns Hopkins 
University, Baltimore, USA}
\author{Robert Szafron} 
\email{szafron@ualberta.ca}
\affiliation{Department of Physics, University of Alberta, 
Edmonton, Canada}

\begin{abstract}
We compute the spin asymmetry  of the muon decay $\mu \to e \bar \nu_e \nu_\mu $ 
through ${\cal O}(\alpha^2)$ in  perturbative QED.
These two-loop  corrections are about 
a factor five (twenty) smaller than the current statistical (systematic) 
uncertainty of the most precise measurement, performed by 
the TWIST collaboration.  We point out that at ${\cal O}(\alpha^2)$ the
asymmetry requires a careful definition due to multi-lepton 
final states and suggest to use familiar QCD techniques to define it
in an infra-red safe way. 
We find that the  TWIST
measurement of the asymmetry is in excellent agreement 
with the Standard Model.
\end{abstract}

\maketitle

\section{Introduction} 

The muon decay is a paradigm for all charged current flavor transformations.
It is a purely leptonic process, $\mu\to e\bar{\nu}_{e}\nu_{\mu}$,
whose properties can be theoretically predicted with very high precision.
Measurements of its lifetime \cite{Tishchenko:2012ie} and distributions
of the daughter electron \cite{TWIST:2011aa,Bueno:2011fq,Bayes:2011zza}
determine fundamental parameters of the Standard Model and probe its
extensions. 

Since muons produced in decays of pions are polarized, the angle $\theta$
between the muon spin direction and the daughter electron momentum can be observed.
The electron distribution in space is
\begin{equation}
\frac{\d\Gamma\left(\mu^{-}\to
    e^{-}\bar{\nu}_{e}\nu_{\mu}\right)}{\d\cos\theta}
=\frac{\Gamma+ \Gamma_0 A \cos\theta}{2},
\label{eq:Theta}\end{equation}
where $A$ is the asymmetry and $\Gamma_{0}= {\mathrm G}_{\mathrm F}^{2}m_{\mu}^{5}/(192\pi^{3})$ 
is the muon decay rate in the massless electron limit,
 and without radiative corrections. ${\mathrm G}_{\mathrm F}$
is the Fermi constant.

The decay rate of an unpolarized muon decay, given by the $\Gamma$-term
in Eq.~\eqref{eq:Theta}, has been extensively studied both theoretically
and experimentally. It was the first decay process of a charged particle
to which one-loop \cite{Behrends:1956mb} and, four decades later,
two-loop \cite{vanRitbergen:1998yd} corrections were computed. Together
with the recent measurement \cite{Webber:2010zf}, these results give
the best value of the Fermi constant ${\mathrm G}_{\mathrm F}$, one of the pillars
of precise electroweak studies. Corrections to more differential quantities 
such as the energy spectrum of electrons, were considered in 
Refs.~\cite{Arbuzov:2002pp,Arbuzov:2002cn,arbuzov2,Anastasiou:2005pn}.

The $A$-term in Eq.~\eqref{eq:Theta} is less well studied, and is the
subject of the present  paper. Since $\cos \theta \sim \vec{ s} \cdot {\vec p}_e$, 
it violates parity and, as such,  it was central in
establishing the structure of the electroweak interaction. Indeed, the
two experiments \cite{garwin57,FriedmanTelegdi57} that confirmed
Madame Wu's discovery of the parity non-conservation \cite{WuParity},
observed the angular asymmetry of the positron distribution in the
antimuon decay. 

Before we describe our calculation in detail, we briefly discuss
the origin of the simplicity of Eq.~\eqref{eq:Theta}, neglecting the
electron mass and radiative corrections. This simple decay pattern is
due to the spin 1/2 of the muon. If we do not observe neutrinos nor the
polarization of the daughter electron, two functions of the electron
energy 
fully describe the decay distribution. They are the
probability amplitudes $\amp_{\pm}$ of the electron
emission along the muon spin, and in the opposite direction. 

\begin{figure}[t]
  \centering
 \includegraphics[width=45mm]{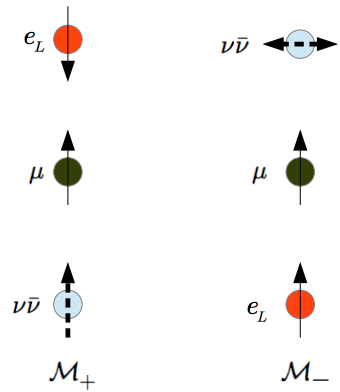}
  \caption{Amplitudes describing the polarized muon decay. The electron
  spin is opposite to its momentum.
  If $e$ is emitted along the muon spin, the projection
  of the total neutrino spin (dotted line) on the $z$ axis must be $+1$; if $e$
  is emitted in the opposite direction, the $\nu\bar\nu$ spin projection vanishes.}
  \label{fig:Mpm}
\end{figure}

Indeed, the probability amplitude for the emission of the electron in
another direction, described by spherical coordinates $\theta$ and
$\phi$ with respect to the muon spin, follows from the spin 1/2
rotation,
\be
\amp\left(\theta,\phi\right)=
\cos\frac{\theta}{2}\amp_{+}
+i\sin\frac{\theta}{2}e^{i\phi}\amp_{-}.
\ee
Since the electron is produced left-handed,
the amplitude $\amp_{+}$ describes the situation when the electron spin
points against the muon spin; thus, the projection of the angular momentum 
carried by neutrinos on the electron momentum should be 
minus one, cf.~Fig.~\ref{fig:Mpm}.
This is easy to arrange when the neutrinos are flying back-to-back, since 
the helicities of $\nu_{\mu}$ and $\bar{\nu}_{e}$ are opposite, as happens when the neutrinos carry most of the energy and 
 the electron little. For the electrons of the highest
energy, $\amp_{+}$ vanishes. 
Conversely,  $\amp_{-}$ describes the configuration when
the electron has the same spin projection as the muon, and the projection
of the neutrinos' angular momentum on the electron direction of motion 
 vanishes. This favors configurations with both neutrinos going in the same direction. 
Relative to $\amp_{-}$, the amplitude $\amp_{+}$  
contains a factor $\sqrt{2\left(1-x\right)}$ (from the Lorentz boost of the
polarization vector of the $\nu\bar\nu$ pair, treated as a 
spin-one particle of mass $m_{\nu\bar\nu} =\sqrt{1-x} \; m_\mu$  where $x = 2E_e/m_\mu$). 
$\amp_{+}$ therefore vanishes for $x=1$.  As a result,
there is a parity violating asymmetry of the electron distribution
with respect to the muon spin, favoring the production of high energy
electrons in the direction counter to the muon spin. Since the muon
decay is suppressed at small electron energies, the asymmetry averaged
over electron energies is negative.

Precise studies of angular effects in the muon decay turned out to be
challenging for both experiment and theory.  Measurements of angular
distributions have been performed ever since the pioneering study
\cite{PhysRevLett.2.56} following the discovery of parity violation.
The results are usually presented in terms of the product of the
degree of the muon polarization $P$ and $\xi$, one of the so-called
Michel-Kinoshita-Sirlin parameters \cite{Beringer:1900zz}. 
It is  related to the decay asymmetry \cite{Beringer:1900zz} 
by $A=|P\xi| A^{\rm th, NLO}$, where $A^{\rm th, NLO}$ is the theoretical  prediction for the 
asymmetry accurate through next-to-leading order in the fine structure 
constant.\footnote{Note that some  
QED corrections beyond NLO were included in the description of 
the electron spectrum in Ref.~\cite{Beringer:1900zz}; however, the included corrections  
do not contribute to the inclusive asymmetry discussed here.}
A deviation of the measured value of $|P\xi|$ from unity may be interpreted as the effect 
of higher-order  QED corrections or effects of physics beyond the Standard Model. 
The current best value is \cite{Bueno:2011fq}
 \begin{equation}
   \label{eq:Pxi}
 |\P \xi|  = 1.00084^{+0.00029 +0.00165}_{-0.00029-0.00063}.
 \end{equation}
where the first error is statistical and the second systematic. 

Since $\alpha/\pi \sim  2 \cdot 10^{-3}$, where $\alpha$ is the fine
structure constant, theoretical 
prediction for  the asymmetry beyond one-loop may be expected to be small. 
Nevertheless, it is interesting to calculate   two-loop effects  for a variety of reasons. 
First, given the definition of $P\xi$,  the two-loop correction to the asymmetry 
is the first QED  effect that  may explain a small deviation of $P \xi$ from $1$. 
Second, 
there is an intrinsic ambiguity in defining the polarization asymmetry
in events with additional electron-positron pairs that appear at NNLO
for the first time.  This ambiguity leads to the infra-red enhancement
of the NNLO QED corrections by $ (\alpha/\pi)^2 \ln m_\mu^2/m_e^2$, so
they might be larger than the naive counting suggests.  Careful
definition of the asymmetry is needed if we use the massless electron
approximation.  We discuss this in detail in Section~\ref{sec:ejets}.
Finally, as we explain below, current computational  technology makes this 
formidable calculation  possible.  This fact is quite impressive since, in general,  
 progress with evaluation QED  corrections to the asymmetry was 
slow. Although the one-loop 
asymmetry was computed in 1958 \cite{Kinoshita:1958ru}, its
dependence on the electron mass was determined only
in 2001 \cite{Arbuzov:2001ui},  five 
years after the two-loop effects  were obtained  for the
lifetime.

The remainder of the paper is organized as follows. In the next
Section we discuss technical details of the computation. In
Section~\ref{sec:ejets} we discuss multi-electron final states and
describe an  infra-red safe definition of the asymmetry.  In
Section~\ref{sec:disc} we provide numerical results for the asymmetry
and discuss their significance for the interpretation of measurements.

\section{Details of the calculation}
\label{sec:det}

Using the simplicity of the differential spectrum,
Eq.~\eqref{eq:Theta}, we obtain the asymmetry $A$  by choosing the 
degree of polarization $P = 1$,
\begin{equation}
\Gamma_0 A=\left.\frac{\d\Gamma}{\d\cos\theta}\right|_{\cos\theta=+1}
  -\left.\frac{\d\Gamma}{\d\cos\theta}\right|_{\cos\theta=-1}.
\end{equation}
Hence, we need to compute the difference of the muon decay rates for
the cases when the muon spin points in the direction of the outgoing
electron and when the muon spin points in the opposite direction.

Calculation of the differential decay rate for the polarized muon
${\rm d} \Gamma$ is performed using the numerical code developed for
the computation of NNLO QCD corrections to semileptonic top and bottom
quark decays~\cite{Brucherseifer:2013iv, Brucherseifer:2013cu}.  In
turn, that calculation was made possible by novel methods developed
for computations of higher-order perturbative corrections in QCD
\cite{Czakon:2010td,Czakon:2011ve,Boughezal:2011jf}, employing a
combination of sector-decomposition and phase-space partitioning to
extract and cancel soft and collinear divergences in a systematic way.

The code developed for the studies of unpolarized quark decays
\cite{Brucherseifer:2013iv, Brucherseifer:2013cu} averages over their
spins.  It has to be modified to deliver the spin asymmetry. We did so
in two ways, obtaining identical results.

First, we recalculated all NNLO amplitudes, keeping the muon spin
quantization axis arbitrary.  Second, the original amplitudes
\cite{Brucherseifer:2013iv, Brucherseifer:2013cu}, also determine
amplitudes for an arbitrary quantization axis.  Indeed, denote by
${\cal }A_{\lambda, \vec n}$ the decay amplitude with the muon spin
quantization axis $\vec {n}$ and helicity $\lambda$. The amplitudes
with  different quantization axes are related by a linear
transformation
\begin{equation}
A_{\lambda_2,\vec n_2}= \sum \limits_{\lambda_1 = \pm} A_{\lambda_1, \vec n_1} \;
\rho_{\lambda_2, \vec n_2 ; \lambda_1,\vec n_1 },
\label{eq:1}
\end{equation}
where the  complex numbers $\rho_{\lambda_2,\vec n_2 ; \lambda_1,\vec
  {n}_1}$ describe a rotation between spinor bases.  Assuming that 
$\vec n_i = (\sin \theta_i \cos \varphi_i, \sin \theta_i \sin \varphi_i,  \cos \theta_i)$, 
we find 
\be
\begin{split} 
& \rho_{+,\vec n_2, +, \vec n_1}  = c_2 c_1 + s_2s_1 e^{i \varphi_{21}},
\\
& \rho_{+,\vec n_2, -, \vec n_1}  =  -c_2s_1+s_2s_1 e^{i \varphi_{21}},
\\
& \rho_{-,\vec n_2, +, \vec n_1}  = -\rho^*_{+,\vec n_2, -, \vec n_1}e^{i \varphi_{21}},
\\
& \rho_{-,\vec n_2, -, \vec n_1}  = \rho^*_{+,\vec n_2, +, \vec n_1} e^{i \varphi_{21}},
\end{split} 
\ee
where $c_i\equiv \cos \theta_i/ 2$, $s_i\equiv
\sin \theta_i/ 2 $, and $\varphi_{21} \equiv \varphi_2 - \varphi_1$.
We use Eq.~(\ref{eq:1})
to translate between the original amplitudes and those where the muon
spin quantization axis is suitable for the calculation of the asymmetry.

\section{Multi-electron final states}
\label{sec:ejets}

At both the leading and the next-to-leading order in $\alpha$, 
muon always decays to a final state with a single electron. But at NNLO,
the final state can have an additional electron-positron pair
$\mu^{-}\rightarrow e^{-}e^{-}e^{+}\bar{\nu_{e}}\nu_{\mu}$.  In the approximation 
when electron mass is neglected, this process is not separately collinear-safe 
since a collinear $e^+e^-$ pair is indistinguishable from a photon. 

Moreover,  multi-electron final states pose a problem for the computation 
of the asymmetry: which of the electrons should  define the muon
quantization axis?  The algorithm that selects the quantization axis 
should be infra-red  
and collinear safe to ensure the cancellation of singularities in the 
final result.

Suppose we decide to choose the direction of the hardest electron in the computation 
of the asymmetry. This choice creates no problem if this electron is produced in the 
hard $\mu \to e \bar \nu \nu$ transition. However, if the hardest electron originates 
from the photon splitting into  a collinear $e^+e^-$ pair, and if its momentum is picked up 
as the direction to compute the asymmetry, the counter-term for this amplitude 
will have the {\it photon} momentum as the reference direction for the asymmetry. 
This counter-term will therefore not cancel with the divergence of the virtual 
correction where there is just one electron in the final state so that its direction 
is automatically taken as the quantization axis for the muon spin. 

Hence, the issue of the definition of the spin asymmetry is subtle.
However, it is similar to infra-red problems encountered in the context
of the quark jets forward-backward asymmetry in perturbative QCD
\cite{Catani:1999nf}.  A full solution depends on experimental
details, including how electrons and photons are operationally
defined.  Unfortunately, such details, and especially a discussion of
multi-electron final states, are absent in Ref.~\cite{Beringer:1900zz}.

To address this issue in a way that is theoretically sound and has a
potential to make a contact with experiment, we decided to define the
spin asymmetry in terms of infra-red and collinear-safe objects,
echoing similar studies of the forward-backward asymmetry in
perturbative QCD \cite{Weinzierl:2006yt}. To this end, for each muon
decay event with an arbitrary final state, we will define a set of
electron and photon jets, and then use the {\it hardest} among the
reconstructed {\it electron jets} as the direction to calculate the
asymmetry.  This is legitimate because
Eq.~(\ref{eq:Theta}) remains valid if we interpret the angle $\theta$
there as the direction of the electron jet rather than the direction
of the electron proper.

The theory of an infra-red safe definition of jet algorithms is well developed 
in QCD (see e.g.~Ref.~\cite{gs}  for a review). However, traditional jet algorithms 
are flavor-blind which is unacceptable for us since we need well-defined ``electron'' jets. 
The required modification  was worked out in \cite{Banfi:2006hf}
and we borrow a suitable jet algorithm  from that paper.  The Durham 
jet algorithm, that allows tracking the jet ``flavor'',  is defined by its distance measure 
that we take to be 
\begin{equation}
\begin{split} 
& y_{ij}^{(F)} = \frac{2(1-\cos\theta_{ij})}{m_\mu^2} \\
& \times \left\lbrace \begin{array}{l}
\max(E_i^2,E_j^2), \text{softer of $i,\, j$ is flavored,} \\ 
\min(E_i^2,E_j^2), \text{softer of $i,\, j$ is flavorless,}
\end{array} \right.
\label{jetA}
\end{split} 
\end{equation}
and by the clustering procedure that we take to be a simple  addition of the four-momenta 
of partons that are re-combined to a jet.  When the measure in Eq.~(\ref{jetA}) 
is applied to an event in the muon decay,  the flavor  of a parton is equated to 
its  electric 
charge and the flavor of a jet is given by the  sum of flavors of its constituents. 
The procedure is iterative: partons are re-combined into a pre-jet if 
a distance $y_{ij}^{(F)}$ between them is smaller than some chosen value $y$ and 
the algorithm continues until no further re-combinations are possible. 

With this modification, the asymmetry is calculated with respect to the direction 
of the  hardest of the 
electron jets, if more than one are reconstructed by the jet algorithm, 
or with respect to the direction of the double-electron jet if both electrons 
end up in a single jet.  In the limit, when the jet resolution parameter vanishes,  
$y \to 0$, the ill-defined no-jet computation of the asymmetry should be 
recovered. This  means that, at order $\alpha^2$, the asymmetry contains $\alpha^2 \ln y$
terms. 

To choose the  jet resolution parameter $y$ in a sensible way, 
we note that a high-energy electron predominantly emits photons  in a
cone of the size $\theta \sim m_e/E_e$ around its direction. We imagine that 
those photons should be treated as part of the electron jet, while photons 
emitted at larger angles should be distinguishable  experimentally. 
Hence, a physics-motivated 
choice of the jet resolution parameter is 
\be
y \sim \frac{\theta^2 E^2}{ m_\mu^2} \sim \frac{m_e^2}{m_\mu^2} \sim 2 \cdot 10^{-5}.
\ee
Note that, for this choice of the resolution parameter, the magnitude of 
$\alpha^2 \ln 1/y$ is similar to that
of  $\alpha^2 \ln (m_\mu^2/m_e^2)$ which would have appeared, had the
mass of the electron been retained.
Although $\ln (10^{5})  \sim 12 \gg 1$, 
a resummation of $\alpha^2 \ln 1/y$ 
corrections  is not needed 
because the QED coupling constant is small, so that $\alpha^2 \ln 1/y \ll 1$ anyway, 
and also because the contribution of 
multi-electron states (the only place where such enhanced corrections appear) 
to the asymmetry is  relatively small. 

\section{Results and discussion}
\label{sec:disc}

Choosing the jet resolution parameter $y = 10^{-5}$, we find the asymmetry 
\begin{equation}
\label{eq:ath}
  A   =A_0 \left[ 1 - 2.9451 \;\bar a
    + 11.2(1) \;\bar a^2 \right],     
\end{equation}
where $A_0 = -1/3$ is the leading order asymmetry, 
$\bar a = \bar \alpha/\pi$ and $\bar \alpha$ is the ${\overline {\rm MS}}$ QED coupling renormalized at the 
scale $\mu = m_\mu$.  It is slightly larger than the canonical fine structure constant, 
$\bar \alpha = 1/135.90$. 
To compare this result with the experimental measurement, 
we take the ratio of $A$  and $A^{\rm th, NLO}$, obtained by truncating 
Eq.~(\ref{eq:ath}) at order ${\cal O}(\bar a)$. We find\footnote{Note that 
inclusion of mass $m_e/m_\mu$ corrections at leading and next-to-leading 
order cannot affect the result for $|P \xi|$ since its deviation from unity 
can only start at ${\cal O}(\alpha ^2)$. Hence, it is consistent to evaluate 
$|P\xi|$ in the massless approximation for two first orders in the expansion 
in the fine structure constant.}
\be
|\P \xi|^{\rm th} = 1.00006(1),
\ee
where the error reflects the sensitivity to $y$, discussed in the next paragraph.
This result is approximately one sigma below  
the measured value of $|P \xi|$, Eq.~(\ref{eq:Pxi}).  The NNLO correction to 
$P \xi$ evaluates to $0.6 \cdot 10^{-4}$; it is therefore 
much smaller than the current experimental precision of $2.9\cdot 
10^{-4}$ (statistical) and $\sim 6 \cdot 10^{-4}$ (systematic). 

\begin{figure}[t]
  \centering
  \includegraphics[width=45mm,angle=-90]{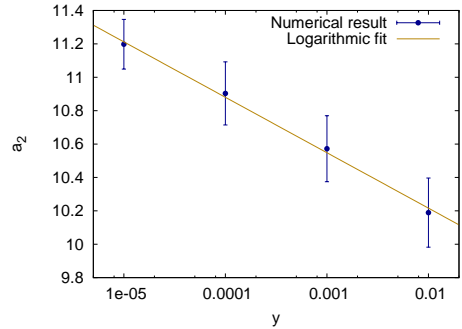}
  \caption{Dependence of the second order relative correction to the asymmetry on the 
jet resolution parameter $y$. The solid line is the fit to the 
function $c_1 + c_2 \ln y$, with $c_1 =9.5(1)$ and $c_2 = -0.14(1) $.}
  \label{fig:jetdep}
\end{figure}

We note that our result for the asymmetry depends on the jet resolution parameter, but 
this dependence is  weak, as we explain below. 
Indeed,  the leading order asymmetry $A_0$ is
independent of $y$.  The NLO coefficient exhibits a linear dependence
on $y$ for small $y$.  The dependence of the NNLO correction to the
asymmetry on the jet resolution parameter for small $y$ can be
approximated by $9.5-0.14 \ln y $, as shown in
Fig.~\ref{fig:jetdep}.  It follows from that figure  that a change in the jet
resolution parameter from $10^{-5}$ to $10^{-2}$, changes the second
order correction to the asymmetry by $10$ percent. As we already
noticed, a choice of $y$ is, in some sense, equivalent to
understanding a correspondence between the measurement setup
\cite{Beringer:1900zz} and the theoretical calculation reported here.  
It follows from Fig.~\ref{fig:jetdep} that this issue
becomes relevant when the precision of the asymmetry
measurement becomes comparable to $ 0.6 \cdot 10^{-5}$, far
from the current level.

The acceptance of the TWIST experiment is fairly complicated; electrons are 
accepted in particular angular and energy regions.
It is therefore interesting to understand 
to what extent the asymmetry depends 
on the electron energy range selected in the experimental analysis. 
To find out, we computed four values of the asymmetry that differ by the cut on the minimal  
energy of the electron jet   $E_{\rm min} < E_{\rm jet}$. 
We consider four values of the minimal jet energy cut, $E_{\rm min} = 10,\;20,\;30,\;40~{\rm MeV}$.
The results are shown in Table~\ref{results_table} for two values of the jet resolution 
parameter.  It is apparent from that Table that corrections strongly depend on the value of $y$ 
and on the jet energy interval. This is the consequence of the  electron energy 
 not being a collinear-safe observable in $m_e \to 0$ limit, so that NNLO corrections 
to the electron energy spectrum contain $\ln^2y$-enhanced terms.\footnote{  
Analogous logarithmic corrections to the asymmetry at NNLO $\alpha^2 \ln^{2,1} m_\mu/m_e$ 
were computed in Refs.\cite{Arbuzov:2002pp,arbuzov2}.} 
As follows from Table~\ref{results_table}, QED  
corrections are particularly large  in case of $y = 10^{-5}$, especially 
in the limit when the cut on the minimal  electron jet energy approaches 
the kinematic boundary. 
These results  suggest 
that the asymmetry depends strongly on the electron (or electron jet) energy. 
This feature will hamper the interpretation of results if improved asymmetry measurements 
become available, unless the asymmetry can be theoretically computed
including the experimental cuts. 
Since muon decay experiments do not use the concept of lepton jets, fully differential computations 
with massive electrons may be needed. This remains an interesting challenge for the future.  

\begin{table*}
\centering
\begin{tabular}{ccccccccccc}
\toprule
& \multicolumn{2}{c}{ $a^{(0)} $} & \multicolumn{2}{c}{$a^{(1)}$ } & \multicolumn{2}{c}{$a^{(2)}$} & \multicolumn{2}{c}{$\delta_{\rm NLO},\%$ } & \multicolumn{2}{c}{$\delta_{\rm NNLO},\%$} \\
$ E_{\rm min}$&$y=10^{-5}$&$y=10^{-2}$&$y=10^{-5}$&$y=10^{-2}$&$y=10^{-5}$&$y=10^{-2}$&$y=10^{-5}$&$y=10^{-2}$
&$y=10^{-5}\;\;$&$y=10^{-2}$
\\ \colrule
$10~{\rm MeV} $  & $1.01$  &  $1.01$ & $-4.01$  & $-3.25$ &  $12.6$ & $11.7$  &  $-0.9$  &  $-0.8$&  $6.9 \cdot 10^{-3}\;\;$  & $6.4 \cdot 10^{-3}$\\ 
$20~{\rm MeV}$   &  $1.05$ &  $1.05$ & $-5.96$  & $-3.73$  & $21.1$&  $13.8$   & $-1.3$&    $-0.8$ & $1.1 \cdot 10^{-2}\;\;$&$7.2 \cdot 10^{-3} $ \\ 
$30~{\rm MeV}$   &  $1.05$ &  $1.05$ & $-9.24$  & $-4.19$  & $49.3$&  $15.9$   & $-2.1$ & $-0.9$ & 
$2.6 \cdot 10^{-2} \;\;$&$8.3 \cdot 10^{-3} $ \\ 
$40~{\rm MeV}$   &  $0.87$&$0.87$  & $-11.78$  & $-3.79$ & $98.4$&  $14.1$   & $-3.2$&$-1.0$ & 
$6.2 \cdot 10^{-2} \;\;$&$8.9 \cdot 10^{-3} $ \\ \botrule
\end{tabular}
\caption{
  Dependence of the asymmetry $\left. A \right|_{y,E_{\min}}
  = - \left(a_0 + a_1 \bar a + a_2 \bar a^2 \right)/3$ on  the minimum 
  accepted jet energy $E_{\min}$, for two values of the jet resolution
  $y=10^{-5}$ and $10^{-2}$.  The last two columns show the relative
  NLO and NNLO effects, in percent. The uncertainty in NNLO coefficients due to numerical 
integration errors is at a few percent level. 
}

\label{results_table}
\end{table*}

It is interesting to compare corrections to the asymmetry and corrections to the 
total rate.  Corrections to the rate, computed in Ref.~\cite{vanRitbergen:1998yd}, read
\begin{equation}
  \label{eq:total}
  \Gamma = \Gamma_0\left[ 1 - 1.81 \; \bar a 
 +6.74 \; \bar a^2
  \right].
\end{equation}
Since the decay rate $\Gamma$ is infra-red finite, Eq.~(\ref{eq:total}) is independent of $y$. 
Comparing Eq.~(\ref{eq:total}) and Eq.~(\ref{eq:ath}), we find that corrections 
to the asymmetry are larger.  However, 
the relative size of subsequent  coefficients in the perturbative series is
comparable in both cases. 

Another source of corrections to the asymmetry and to the total rate are  
the electron mass effects ${\cal O}(m_e / m_\mu)$ and 
it  is interesting to compare them with the size of NNLO 
QED corrections, for  both the decay rate and the asymmetry. 
The correction of order $\order{\alpha^2}$ to the total decay
width 
is less important
than the  effect of the electron mass in the lowest order decay rate. 
In the asymmetry, the electron mass effect  is practically
negligible, suppressed by 
additional two powers of $m_e/m_\mu$  \cite{Arbuzov:2001ui}.  Thus, for the asymmetry, the radiative corrections are
more important than the electron mass effects.

Should future tests of the $V-A$ structure of the muon decay
interaction be undertaken, they will not be encumbered with large
radiative corrections.  However, if the experimental precision reaches
the size of the two-loop effects, one will have to carefully match the
details of theoretical computations with the experimental setup.

Our approach can also be used to compute  the NLO QED 
corrections to the radiative decay of a polarized muon.  Such computation 
is  important for controlling the background to the search of $\mu\to
e\gamma$, and the need for such a correction has recently been pointed
out \cite{Adam:2013gfn}.  The main modification necessary in that
study will be the optimal parametrization of the phase space,
convenient for the rather extreme kinematics interesting from the
experimental point of view.

\noindent
{\bf Acknowledgments:}  We would like to thank S.~Catani for 
comments about infra-red safety of the spin asymmetry. 
R.S. is grateful to the Particle Theory Group at Johns 
Hopkins University for the hospitality extended to him during the work on this 
paper.  The work of K.M.  and F.C.  
is partially supported by US NSF under grants PHY-1214000. 
A.C., R.S., and Y.L.~are supported by Science and Engineering Research
Canada (NSERC).


\end{document}